\documentclass[draft]{agujournal2019}
\usepackage{url} 
\usepackage{lineno}
\usepackage{amsmath}
\usepackage{amssymb}
\usepackage{booktabs}
\usepackage[inline]{trackchanges} 
\usepackage{soul}
\usepackage{placeins}
\usepackage{float}
\usepackage{dsfont}

\journalname{Space Weather}

\begin{document}

\title{C-SWIM: A Coupled Space Weather Impact Model for Satellite Fleet Vulnerability and Economic Loss Under a 1-in-100-Year Solar Energetic Particle Event}

%
%

\authors{D. Bor\affil{1,2}, E. J. Oughton\affil{1,2}, R. S. Weigel\affil{1,2}, R. Yang\affil{1}, T. Clower\affil{3}, M. J. Wiltberger\affil{4}, R. Linares\affil{5}}

\affiliation{1}{Department of Geography and Geoinformation Sciences, George Mason University, Fairfax, VA}
\affiliation{2}{Space Weather Lab, George Mason University, Fairfax, VA}
\affiliation{3}{Schar School of Policy and Government, George Mason University, Fairfax, VA}
\affiliation{4}{NSF NCAR/HAO, Boulder, CO}
\affiliation{5}{Massachusetts Institute of Technology, Cambridge, MA}

\correspondingauthor{E. J. Oughton}{eoughton@gmu.edu}






\begin{keypoints}
\item A 1-in-100-year SEP event places approximately 100 of about 10,650 US operational satellites in the Critical failure risk class
\item Expected capital loss totals 5.2 billion dollars and is concentrated in high-altitude LEO and HEO assets rather than GEO
\item Daily economic impact ranges from 70 million to 1.3 billion dollars across three failure scenarios with Earth observation most disrupted
\end{keypoints}

\begin{abstract}
Modern economies depend critically on satellite infrastructure, yet the aggregate economic consequences of extreme solar energetic particle (SEP) events have not been rigorously assessed. This study develops an integrated framework linking SEP hazard characterization, dynamic geomagnetic cutoff rigidity modeling, radiation dose transport, and fleet-wide failure probability estimation to macroeconomic impact analysis. Using extreme-value analysis of 160 SEP events over 27.4 years (1996--2025), failure probability is estimated for ${\sim}$10,650 US operational satellites under orbital regime-dependent shielding assumptions. The assessment reveals that ${\sim}$100 satellites (1.0\%) are at Critical risk, concentrated in high-altitude low Earth orbit and highly elliptical orbit, while medium Earth orbit and geosynchronous orbit satellites fall in the Negligible class ($P_{\mathrm{fail}} < 10^{-9}$) under the assumed radiation-hardened components and shielding. The expected capital loss across the ${\sim}$\$254B fleet totals ${\sim}$\$5.2B. Three failure scenarios, expanding from Critical satellites only ($P_{\mathrm{fail}} > 10^{-2}$), to Critical and Elevated ($P_{\mathrm{fail}} > 10^{-3}$), and to all satellites with non-negligible risk ($P_{\mathrm{fail}} > 10^{-6}$), yield daily economic impacts of ${\sim}$\$70M, ${\sim}$\$270M, and ${\sim}$\$1.3B, respectively. Earth observation suffers up to 95.6\% capacity loss in the worst case, while military services experience 16.1--20.4\% disruption across scenarios. Results are first-order estimates: hardware failure counts are conservative because only total ionizing dose is modeled, and daily economic impacts represent upper bounds because operator response and recovery are not included.
\end{abstract}

\section*{Plain Language Summary}
A once-in-a-century solar storm would likely damage about 100 US satellites valued at \$22 billion, with an expected loss of roughly \$5 billion after accounting for the chance each satellite actually fails. These satellites are mostly in high-altitude or elliptical orbits, where they already endure heavy exposure to Earth's radiation belts. GPS satellites are well protected and remain fully operational under the likely and moderate scenarios, while commercial communications experience only 0.8 to 7.4 percent disruption thanks to constellation redundancy. Earth observation and military surveillance are the most affected, contributing to total daily economic losses of \$70 million to \$1.3 billion across all sectors.

\section{Introduction}\label{sec:introduction}

As Solar Cycle 25 peaks, there is an increased risk of solar activity affecting technological infrastructure. A notable example is the May 2024 Gannon storm, where multiple X-class flares and Earth-directed coronal mass ejections (CMEs) from active region AR3664 triggered significant geomagnetic disturbances, with auroral activity observed as far south as 21$^\circ$N latitude \cite{parker_satellite_2024, hajra_interplanetary_2024, pierrard_mothers_2024}. During such events, elevated solar energetic particle (SEP) fluxes pose significant risks to spacecraft microelectronics, potentially causing single-event effects (SEEs) and degrading solar panel efficiency.

SEPs, primarily energetic protons, heavy ions, and electrons accelerated during CMEs and solar flares, pose significant hazards to space-based systems \cite{petersen2007radiation, koga2007effects, pulkkinen_space_2007}. When these particles penetrate Earth's magnetosphere, they can cause variations in atmospheric chemistry and dynamics, adversely affecting space-based and ground-based assets \cite{baker_satellite_2001, schwenn_space_2006, whitman_review_2023}. Satellite vulnerability to SEP exposure depends on geomagnetic shielding, which varies with orbital position and storm-time magnetospheric field configuration. Particle rigidity, defined as momentum per unit charge (GV), determines whether a proton can overcome the local magnetic barrier and reach a given orbital location. Geomagnetic cutoff rigidity $R_c$ is the minimum rigidity required for this condition to be satisfied \cite{smart2006geomagnetic}. Under quiet conditions, Earth's magnetic field deflects particles below $R_c$, providing protection at equatorial latitudes. However, during intense geomagnetic storms, this shielding erodes, allowing SEPs to access previously inaccessible orbital regions \cite{smart_review_2005}.

While the physics and engineering understanding of space weather effects has advanced considerably, the linkage between physical hazards and economic consequences remains underdeveloped. Extensive research exists on radiation effects in individual electronic components \cite{petersen_seu_1998, hansen_review_2024}, and studies have examined correlations between satellite anomalies and geomagnetic indices \cite{choi_analysis_2011}. Yet fleet-wide vulnerability assessments that connect physical SEP exposure to economic outcomes are notably absent. Economic risk quantification of space weather has received growing attention, including studies on power grid and geomagnetically induced currents (GIC) impacts \cite{oughton_quantifying_2017, eastwood_economic_2017}, global supply chain disruptions \cite{schulte_how_2014}, sector-specific impact surveys \cite{ishii_space_2021}, and Global Positioning System (GPS) outage costs \cite{oconnor_gps_economic_2019}. However, these estimates typically rely on scenario-based approaches, without explicit uncertainty propagation or direct coupling to probabilistic geophysical hazard models. Recent work has demonstrated coupled physics-engineering-economic frameworks for power grid infrastructure \cite{oughton2024major}, but no equivalent framework exists for satellite systems, motivating the present study. Extending such approaches to satellites requires addressing unique challenges, including characterizing the orbital radiation environment using magnetospheric and particle-tracing models, propagating satellite-fleet trajectories across diverse orbital regimes, and mapping service disruptions to sectoral economic impacts.

Thus, this study develops and applies an integrated, multi-layered modeling framework linking physics-based space weather hazard analysis to macroeconomic impact estimation. It couples dynamic geomagnetic shielding modeling, radiation dose transport through spacecraft shielding, and vulnerability assessment of satellite electronics under orbital-regime-dependent shielding and component-hardness assumptions. Losses from space weather-related failures are then estimated using macroeconomic input-output (IO) analysis. Using extreme-value analysis of 27.4 years of SEP observations (1996--2025), we derive 1-in-100-year storm scenarios and predict fleet-wide vulnerability for 10,650 US operational satellites. Central to this framework is the role of geomagnetic cutoff rigidity erosion, which, during extreme storms, exposes previously shielded orbital regions to damaging particle fluxes. The resulting SEP access, combined with accumulated trapped radiation dose and extreme-event fluence, determines total dose and failure probability across the fleet. This paper addresses three research questions:

\begin{description}
    \item[1.] What is the SEP exposure of US operational satellites during extreme geomagnetic storms, as characterized by dynamic cutoff rigidity erosion and particle access fractions?
    \item[2.] What is the failure probability of the US operational satellite fleet under extreme SEP events, accounting for accumulated trapped radiation dose and orbital-regime-dependent shielding?
    \item[3.] What are the economic impacts of SEP-induced satellite service disruptions, including capital asset valuation and downstream impacts on the broader US economy?
\end{description}
\noindent

This paper is organized as follows: Section~\ref{sec:background} provides background on radiation hazards and satellite exposure. Section~\ref{sec:method} presents the methodology, Section~\ref{sec:results} presents the results, and Section~\ref{sec:discussion} interprets the findings with respect to the research questions.
\section{Background}\label{sec:background}

\subsection{Solar energetic particles and magnetospheric response}

SEP events originate from two distinct acceleration processes, namely CME-driven shock acceleration and magnetic reconnection in solar flares. CME-driven interplanetary shocks produce large, long-duration gradual events with high proton fluences \cite{desai_large_2016, ryan_solar_2000}. Magnetic reconnection in solar flares produces shorter, electron-rich impulsive events that are generally lower in intensity \cite{reames_solar_1995, lin_relationship_2005}. Gradual SEP events are of particular concern for spacecraft radiation exposure, as their high-energy proton fluences can span several orders of magnitude depending on event intensity and solar cycle phase \cite{desai_large_2016, baker_solar_2008}. The extent to which these particles penetrate the magnetosphere is quantified through cutoff rigidity. Cutoff-based characterization can differ significantly between low-altitude and high-altitude measurements \cite{obrien_solar_2018}. During geomagnetic storms, the erosion of this shielding expands SEP access to lower orbital altitudes, elevating proton flux levels and cumulative dose rates that directly affect SEE rates and total ionizing dose budgets for low Earth orbit (LEO) missions \cite{girgis_geomagnetic_2023}.

\subsection{Radiation effects on spacecraft}

Space weather impacts satellite operations through three primary mechanisms: internal and surface charging, SEEs, and cumulative degradation. Charging is commonly associated with disturbed electron environments, while SEEs represent a key pathway for immediate disruptions when energetic particles interact with sensitive microelectronics. In practice, SEEs can range from transient upsets to latch-ups, which are self-sustaining high-current states in semiconductor devices that require ground intervention to reset and can trigger broader subsystem interruptions \cite{rycroft_electromagnetic_2017}. In addition to the traditional emphasis on higher-energy particles, direct ionization by lower-energy protons can also contribute materially to upset rates in modern deep-submicron technologies, with implications for rate prediction and hardness assurance \cite{sierawski_impact_2009}. 

Beyond immediate anomalies, cumulative radiation effects drive economic loss through progressive degradation. Displacement damage and total ionizing dose (TID) reduce solar array efficiency and degrade electronic components over time, shortening a satellite's effective service life \cite{barnaby_total-ionizing-dose_2006, hands_radiation_2018}. At the fleet level, analyses of geostationary Earth orbit anomaly archives and long telemetry records show that anomaly occurrence varies with geomagnetic activity and energetic particle conditions, consistent with a mixture of charging- and radiation-driven contributors \cite{choi_analysis_2011, lohmeyer_space_2013}. Recent work also emphasizes improved monitoring of SEE modes such as latch-up and associated functional interrupts in spacecraft hardware \cite{mattos_investigation_2024}.

\subsection{Geomagnetic shielding and orbital exposure}

The near-Earth radiation environment is structured by the Van Allen belts, with the inner belt extending from the upper atmosphere to $\sim$10,000~km and the outer belt spanning $\sim$3--6~Earth radii ($R_E$), producing altitude- and orbit-dependent particle exposures \cite{lanzerotti_space_2017}. For SEPs, access to low- and mid-latitude near-Earth space is largely controlled by geomagnetic shielding, often quantified by cutoff rigidity \cite{freeman_specifying_2001,smart_fifty_2009}. During geomagnetic storms, this shielding weakens as the magnetosphere is compressed and storm-time current systems intensify, thereby allowing SEPs to penetrate regions typically inaccessible. Case studies show that rigidity cutoff suppression can be rapid and substantial, reaching $\sim$1--1.8~GV in major storms and varying on short timescales in response to changes in solar wind dynamic pressure, interplanetary magnetic field (IMF), and the ring current \cite{kress_solar_2010, danilova_mapping_2019, kichigin_variations_2018, adriani_pamelas_2016}. 

The storm-time cutoff response is also structured in local time. Observations indicate that dayside and nightside cutoff boundaries can shift in opposite directions during storm main phases, producing strong day--night asymmetries that fixed latitude thresholds cannot represent \cite{nesse_tyssoy_cutoff_2015, chu_geomagnetic_2016}. Because cutoff boundaries are frequently used as practical limits on the impact of SEPs, it is important to assess how well they match observations. Riometer-based evaluations show that fixed cutoffs (for example, 60$^\circ$ geomagnetic latitude) can systematically misestimate the impact area for lower-energy protons \cite{heino_observational_2020}. 


\subsection{Policy context and economic estimates}\label{sec:policy}

The broader macroeconomic impacts of space weather are substantial. Extreme space weather events associated with CMEs and solar flares could cause daily losses of \$4--7~billion across the power, satellite, GPS, and aviation sectors \cite{abt_associates_social_2017}. Moreover, the National Oceanic and Atmospheric Administration (NOAA) estimates that space weather forecasting prevents \$111~million to \$27~billion in avoided losses, depending on storm severity \cite{noaa_ergs_2021}. 

Recognition of these economic risks is not new. The National Space Weather Program, established in 1995 under the Office of the Federal Coordinator for Meteorological Services and Supporting Research, represented an early multi-agency effort to enhance national preparedness \cite{robinson2001us}. These concerns intensified in 2008, when the National Academies convened a workshop warning of a potential ``space weather Katrina'' \cite{nrc_severe_2009}. On the policy front, the National Aeronautics and Space Administration (NASA) Authorization Act of 2010 mandated interagency coordination for space weather preparedness \cite{usgov_nasa_authorization_2010}. More recently, the 2019 National Space Weather Strategy established resilience objectives across critical infrastructure sectors \cite{ostp_national_2019}. 

Translating these recognized risks into quantitative impact estimates for satellite systems remains an open challenge. The following section describes our methodology for quantifying satellite fleet exposure to SEP hazards, scenario analysis of satellite maloperation, and cascading the resulting economic impact.

\section{Method}\label{sec:method}

\begin{figure}[H]
    \centering
    \includegraphics[width=\textwidth]{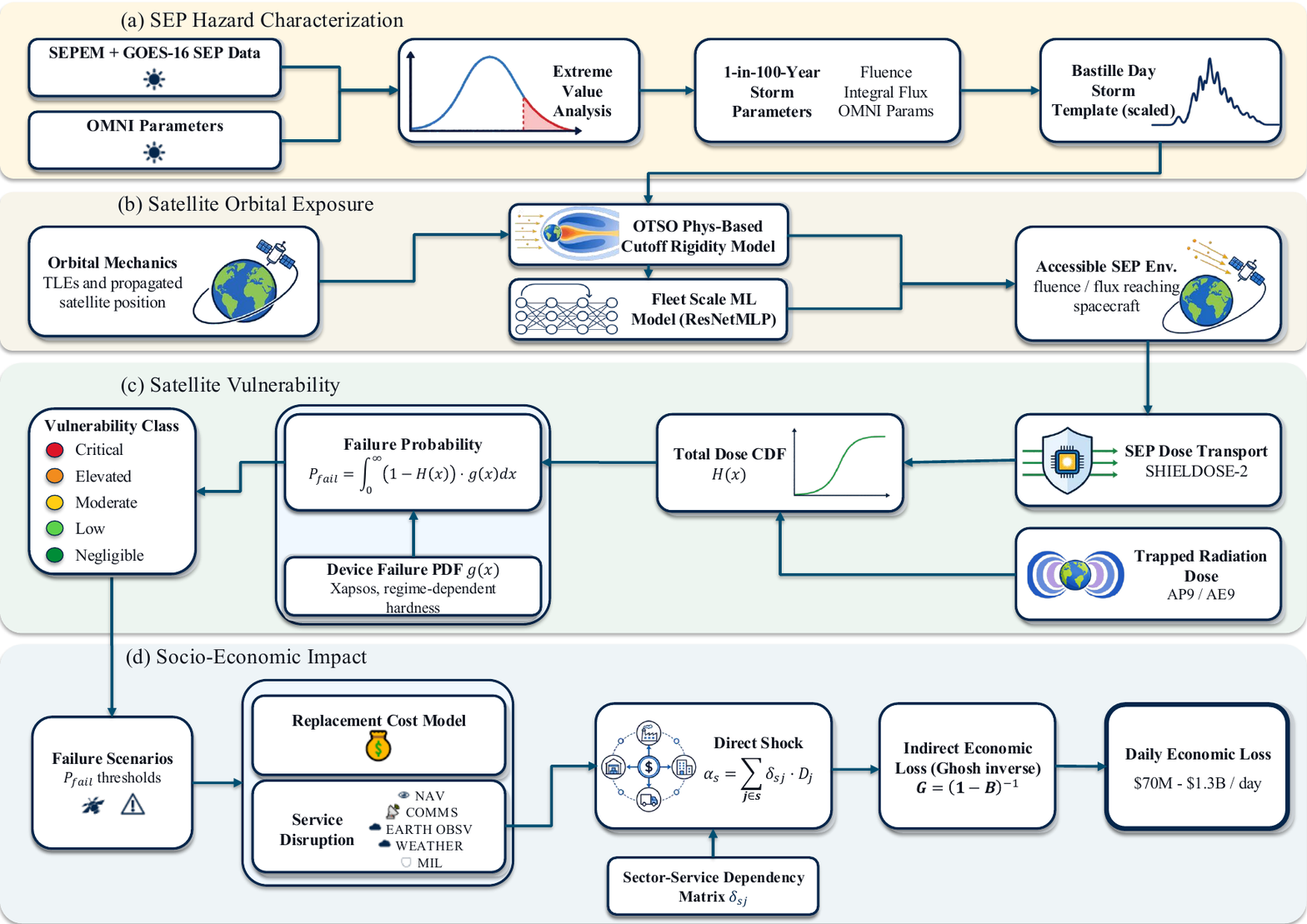}
    \caption{\textbf{Integrated C-SWIM framework for SEP-driven satellite vulnerability and economic impact assessment.} The four panels correspond to the four method stages: (a) SEP hazard characterization, (b) satellite orbital exposure, (c) satellite vulnerability, and (d) socio-economic impact.}
    \label{fig:framework}
\end{figure}

Figure~\ref{fig:framework} illustrates the C-SWIM framework for assessing radiation vulnerability of the US operational satellite fleet under a 1-in-100-year SEP event. The framework first characterizes the SEP hazard from 27.4 years of proton flux observations (Section~\ref{method:sep-hazard}). Geomagnetic shielding along each satellite trajectory is then computed using particle trajectory tracing, with a machine-learning surrogate enabling fleet-scale inference (Section~\ref{method:cutoff}). The resulting particle access fractions are combined with extreme-event fluence and accumulated trapped radiation dose to obtain the total dose environment for each satellite (Section~\ref{method:vulnerability}). Failure probabilities derived from this dose environment are translated into economic impact through scenario-based input-output analysis (Section~\ref{method:economic}).

\newpage

\subsection{SEP hazard characterization}\label{method:sep-hazard}

Integral proton flux measurements spanning 1996--2025 are derived from two complementary sources, with the first detected SEP event occurring in November 1997, yielding an effective observation window of 27.4 years. Following the NOAA Space Weather Prediction Center (SWPC) criterion, an SEP event is defined as a period during which integral proton flux $J_{>10~\mathrm{MeV}} > 10$~pfu, where 1~pfu $= 1$~proton~cm$^{-2}$~s$^{-1}$~sr$^{-1}$ \cite{noaa_space_weather_scales_2026}. Solar Energetic Particle Environment Modelling (SEPEM) Reference Data Set v3.2 provides quality-controlled integral proton flux at 5-minute resolution for the period 1974--2017 \cite{crosby_sepem_2015}. For coverage from November 2020 onward, we used the Geostationary Operational Environmental Satellite 16 (GOES-16) Solar and Galactic Proton Sensor data \cite{noaa_goes16_sgps}. The intervening period 2018--2020 coincides with the Solar Cycle 25 minimum, during which GOES-15 Space Environment Monitor (SEM) observations confirmed no flux exceedances above the 10~pfu threshold \cite{noaa_goes15_sem}.

A total of 160 SEP events are identified across the observation period (128 from SEPEM and 32 from GOES-16). For each event, both peak integral flux $J_{\max,>E}$ and event-integrated fluence
\begin{equation}
    \Phi(>E) = \pi \int_{t_{\mathrm{start}}}^{t_{\mathrm{end}}} J_{>E}(t)\,dt
    \label{eq:fluence}
\end{equation}
\noindent
are extracted at energy thresholds $E = 5$, 10, 30, 50, and 100~MeV, where $t_{\mathrm{start}}$ and $t_{\mathrm{end}}$ are the event onset and end times. The factor of $\pi$ converts from directional to omnidirectional fluence under the assumption of isotropic incidence on a planar shield. For each event, peak values of the associated interplanetary and geomagnetic conditions are also extracted to characterize the storm-time magnetospheric configuration used in the cutoff rigidity modeling of Section~\ref{method:cutoff}, with the full parameter set listed in Table S8 of the Supporting Information.

To estimate the magnitude of a 1-in-100-year event, we fit a statistical tail model to the most extreme observed values in each variable. Specifically, peak-over-threshold (POT) extreme-value analysis is applied to flux maxima, event fluences and storm driver peaks, following established space-weather climatology methods \cite{lucas_geoelectric_2020}. Exceedances above the threshold $u$ are modeled using the generalized Pareto distribution (GPD) \cite{castillo_fitting_1997}:
\begin{equation}
    F(x) = 1 - \left(1 + \frac{\xi(x-u)}{\sigma}\right)^{-1/\xi}, \qquad x > u,
    \label{eq:gpd}
\end{equation}
\noindent
where $F(x)$ is the cumulative distribution function (CDF), $\sigma > 0$ is the scale parameter, and $\xi$ is the shape parameter. Given the limited number of exceedances ($n_{\mathrm{exc}} = 19$ per variable, corresponding to the top 12\% of the 160-event catalog), parameters are estimated using the method of moments (MoM), which provides more stable estimates than maximum likelihood estimation for small samples. The MoM estimators are:
\begin{equation}
    \hat{\xi} = \frac{1}{2}\left(1 - \frac{\bar{x}^2}{s^2}\right), \qquad
    \hat{\sigma} = \bar{x}(1 - \hat{\xi})
    \label{eq:mom}
\end{equation}
\noindent
where $\bar{x}$ and $s^2$ are the sample mean and variance of the exceedances. The $T$-year return level $x_T$ is computed as:
\begin{equation}
    x_T = u + \frac{\sigma}{\xi}\left[(\lambda T)^{\xi} - 1\right]
    \label{eq:return_level}
\end{equation}
\noindent
where $T$ is the return period in years and $\lambda$ is the mean annual rate of exceedances. Parameter uncertainty is quantified using a joint bootstrap procedure ($N_{bs}=1000$ iterations) that resamples events with replacement, preserving inter-variable correlations among solar wind drivers, peak fluxes, and event fluences. The resulting GPD shape parameters for the fluence variables range from $\xi = -0.05$ ($>$5~MeV, bounded tail) to $\xi = +0.09$ ($>$100~MeV, heavy tail), indicating that higher-energy channels exhibit greater variability in extreme event magnitudes.

\begin{figure}[H]
    \centering
    \includegraphics[width=\textwidth]{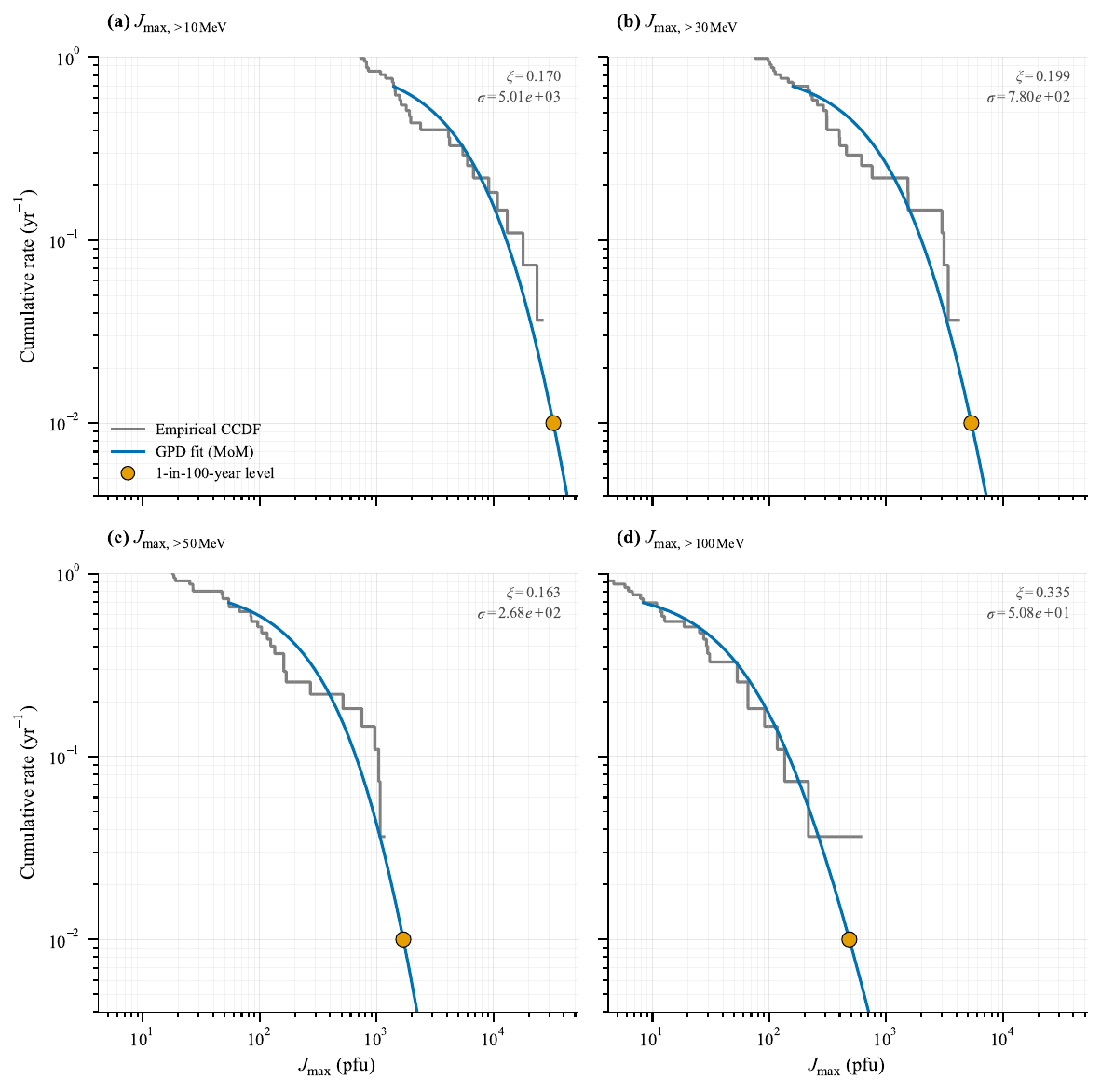}
    \caption{\textbf{Complementary cumulative distribution functions for SEP peak flux.} Empirical CCDFs and fitted GPD models for peak integral proton flux $J_{\max,>E}$ at energy thresholds of 10, 30, 50, and 100~MeV. The y-axis gives the mean number of events per year exceeding a given flux level. The orange dot marks the 1-in-100-year return level. GPD shape ($\xi$) and scale ($\sigma$) parameters estimated by the method of moments are annotated in each panel.}
    \label{fig:ccdf}
\end{figure}

The fit quality is assessed through the complementary cumulative distribution function (CCDF), which gives the mean annual rate of events exceeding a given value for each fitted variable. Figure~\ref{fig:ccdf} shows the empirical and fitted CCDFs for peak flux at each energy channel, with the 1-in-100-year return level marked. Corresponding CCDFs for event-integrated fluence and interplanetary drivers are provided in Figures S2 and S3 of the Supporting Information.

\begin{figure}[H]
    \centering
    \includegraphics[width=\textwidth]{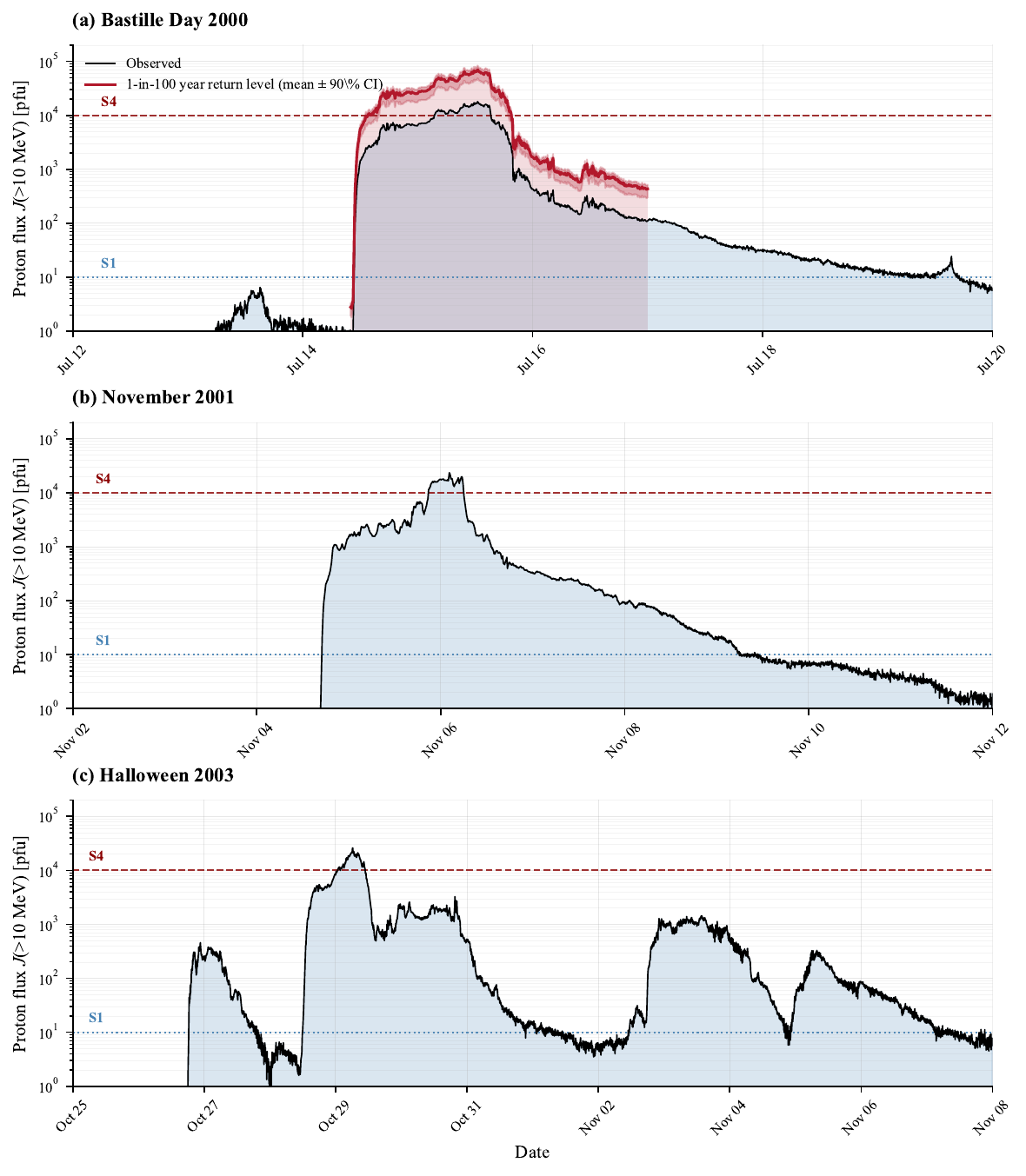}
    \caption{Flux profiles of three major historical solar energetic proton events. The Bastille Day 2000 event (panel a) serves as the baseline template, with the 1-in-100-year return-level scenarios overlaid in red, showing the scaled flux amplitude derived from generalized Pareto distribution extrapolation. Horizontal lines indicate NOAA Solar Radiation Storm thresholds \cite{noaa_space_weather_scales_2026}. S1 (10~pfu) marks the onset of a radiation storm, and S4 (10$^4$~pfu) denotes a severe event with significant risk to satellite electronics, polar-route aviation, and high-frequency radio communications at high latitudes. The observed peak intensity is $J_{>10} = 17{,}900$~pfu, while the 1-in-100-year scenarios yield peak values of $1.7$--$3.8 \times 10^4$~pfu (Table S8 in Supporting Information).}
    \label{fig:baselines}
\end{figure}

The Bastille Day 2000 event (July 14--16) serves as the baseline storm template due to its well-characterized single-peaked flux profile and sustained geomagnetic disturbance, as shown in Figure~\ref{fig:baselines}. A 72-hour driver window centered on the Dst minimum provides the temporal structure for geomagnetic scaling at 5-minute cadence, while the SEP flux window spans the full event duration ($\sim$62~hours). To construct design reference events, the baseline amplitude is linearly scaled:
\begin{equation}
    X_{\mathrm{scenario}}(t) = X_{\mathrm{baseline}}(t) \cdot \frac{x_T}{X_{\mathrm{baseline,peak}}}
    \label{eq:scaling}
\end{equation}
\noindent
where $X_{\mathrm{scenario}}(t)$ is the scaled storm time series at time $t$, $X_{\mathrm{baseline}}(t)$ is the baseline template time series, and $X_{\mathrm{baseline,peak}}$ is the corresponding extremum in the Bastille Day template. The resulting 1-in-100-year scaled storm parameters are summarized in Table S8 of the Supporting Information.

\newpage

\newpage
\subsection{Orbital propagation and satellite population}\label{method:propagation}

To evaluate SEP exposure across the satellite fleet, orbital positions must be determined at every timestep during the storm scenario. Satellite orbital elements are obtained from Space-Track (\url{https://www.space-track.org}) and supplemented with catalog data from Jonathan's Space Report (\url{https://www.planet4589.org}). Satellite trajectories are propagated using methods consistent with the available orbital descriptors. For satellites with two-line element sets (TLEs), we use the Simplified General Perturbations 4/Simplified Deep Space Perturbations 4 (SGP4/SDP4) family of analytical propagators, which model secular and periodic perturbations due to Earth's oblateness, atmospheric drag via the ballistic coefficient $B^*$, and deep-space solar--lunar effects for high-period objects \cite{vallado_revisiting_2006}. For satellites without TLEs, trajectories are generated via numerical integration of the equations of motion with the dominant $J_2$ zonal harmonic (Earth oblateness) perturbation retained \cite{montenbruck_satellite_2000}. Initial state vectors are constructed from the catalog elements available for each satellite, namely perigee altitude, apogee altitude, and inclination, under the convention that the satellite begins at perigee. The full equations of motion and the initial-condition construction are provided in Supporting Information Text S3.

Satellite trajectories are propagated over a 72-hour window with regime-dependent cadence: 60~s for LEO and highly elliptical orbit (HEO), 120~s for medium Earth orbit (MEO), and 300~s for geosynchronous orbit (GEO). Orbital regimes are assigned using period and altitude. GEO is defined by orbital period of $1436 \pm 30$~min and eccentricity $e<0.1$. HEO is defined by $e>0.25$ with apogee above 20{,}000~km. MEO covers perigee $>2{,}000$~km excluding GEO and HEO. LEO covers perigee $\leq 2{,}000$~km. The Bastille Day 2000 event serves as a temporal template for the geomagnetic field profile of the 1-in-100-year scenario, applied to the current satellite fleet independent of the original event date, and the 72-hour trajectory window is constructed to match its duration. Each satellite position along its orbit is paired with the corresponding storm-time field state at each timestep for cutoff rigidity computation in Section~\ref{method:cutoff}. For satellites with TLEs, propagation begins from the TLE epoch, and a fallback epoch of January~15, 2025, is used for the remainder.

\begin{figure}[H]
    \centering
    \includegraphics[width=\textwidth]{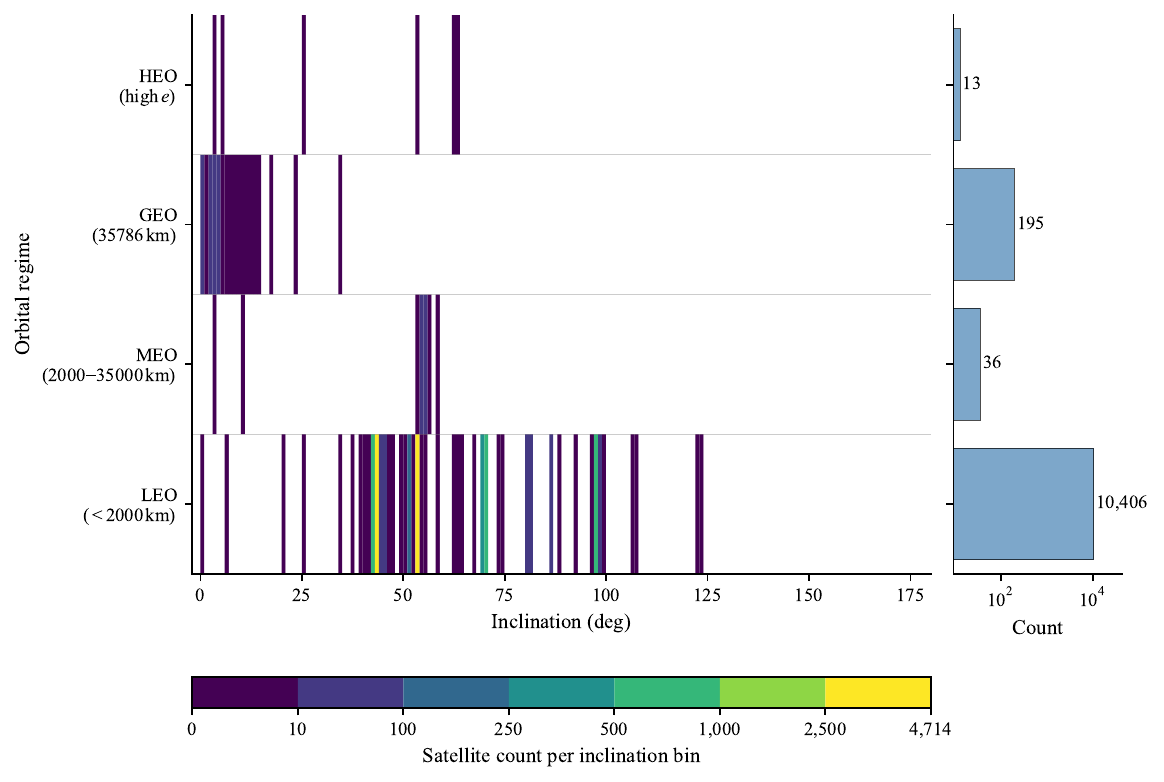}
    \caption{Distribution of US active payload satellites by orbital regime and inclination. Left panel: 2D histogram showing satellite count as a function of inclination for each regime. Right panel: marginal counts per regime (log scale). LEO dominates the population with strong clustering at sun-synchronous ($97^\circ$) and constellation-specific inclinations ($53^\circ$, $52^\circ$).}
    \label{fig:satellite_orbits}
\end{figure}

Figure~\ref{fig:satellite_orbits} illustrates the distribution of the US active payloads across orbital regimes. The majority of satellites occupy LEO, with pronounced inclination clustering near $53^\circ$ (Starlink constellation), $97^\circ$ (sun-synchronous orbits), and $52^\circ$ (International Space Station (ISS)-compatible). The MEO population is dominated by navigation constellations at $55^\circ$ inclination (GPS). GEO satellites have near-zero inclination as expected for equatorial orbits.

\subsection{Geomagnetic cutoff rigidity modeling}\label{method:cutoff}

Having characterized the SEP hazard in Section~\ref{method:sep-hazard} and established satellite orbital positions in Section~\ref{method:propagation}, we next determine which satellites are physically exposed to the energetic proton flux. Geomagnetic shielding during the storm is quantified through the cutoff rigidity $R_c$, which is determined by tracing test proton trajectories backward through the magnetosphere and identifying the minimum rigidity required to reach a given orbital location \cite{smart_fifty_2009}. For protons with kinetic energy $E_k$, rigidity is obtained from
\begin{equation}
    R = \frac{\sqrt{E_k(E_k + 2m_pc^2)}}{qc}
    \label{eq:rigidity}
\end{equation}
\noindent
where $R$ is magnetic rigidity, $m_pc^2 \approx 938.27$~MeV is the proton rest energy, $q$ is the particle charge, and $c$ is the speed of light. Representative conversions yield $R \approx 0.137$~GV for $E_k = 10$~MeV, $R \approx 0.239$~GV for $E_k = 30$~MeV, and $R \approx 0.444$~GV for $E_k = 100$~MeV. By comparison, the quiet-time St\"ormer vertical cutoff rigidity at Earth's surface varies with invariant magnetic latitude $\lambda$ approximately as $R_c \approx 14.9 \cos^4 \lambda$~GV \cite{smart_fifty_2009}, yielding $14.9$~GV at the magnetic equator, $3.7$~GV at $\lambda = 45^\circ$, and $0.9$~GV at $\lambda = 60^\circ$. The 10--100~MeV proton rigidities therefore fall below the cutoff everywhere except at high magnetic latitudes during quiet conditions, with the accessible region expanding equatorward as the cutoff erodes during geomagnetic storms.

Cutoff rigidities are computed using the Oulu Open-source geomagneToSphere propagation tool (OTSO) \cite{larsen_otso_2023}, which numerically integrates charged particle trajectories through the magnetosphere by solving the relativistic Lorentz force equation,
\begin{equation}
    \frac{d\mathbf{p}}{dt} = q (\mathbf{v} \times \mathbf{B}),
    \label{eq:lorentz}
\end{equation}
\noindent
where $\mathbf{p}$ is relativistic momentum, $\mathbf{v}$ is particle velocity, and $\mathbf{B}$ is the magnetic field vector.

The spatiotemporal magnetic field is given by $\mathbf{B}(\mathbf{r},t)=\mathbf{B}_{\mathrm{int}}+\mathbf{B}_{\mathrm{ext}}$, where $\mathbf{r}$ is the position, $\mathbf{B}_{\mathrm{int}}$ is the internal geomagnetic field represented by the International Geomagnetic Reference Field (IGRF-13) spherical harmonic expansion \cite{alken_international_2021}, and $\mathbf{B}_{\mathrm{ext}}$ is the external magnetospheric field represented by the Tsyganenko 2001 (T01) semi-empirical model \cite{tsyganenko_model_2002}. We use T01 rather than later Tsyganenko models (e.g., TS04/TS05) because our statistical driver set includes the $G_1$--$G_3$ coupling functions required by T01 but does not provide the high-cadence Qin--Denton parameters ($W_1$--$W_6$) needed by newer formulations. T01 parameterizes storm-time magnetospheric currents using solar wind dynamic pressure $P_{\mathrm{dyn}}$, Dst, and IMF components ($B_y$, $B_z$).

Direct tracing for the full satellite catalog is computationally expensive. We therefore train a machine-learning (ML) model on OTSO outputs to enable rapid prediction of $R_c$ for fleet-wide assessment. The ML model is a residual multilayer perceptron (ResNetMLP) \cite{he_deep_2016} with 15 input features spanning orbital parameters, magnetic coordinates, and interpolated storm drivers, listed in full in Supporting Information Text S5. Across the held-out test partition, which covers cutoff rigidities from near zero in the polar caps to approximately $15$~GV at the geomagnetic equator, the model achieves a coefficient of determination of $0.9972$. On an independent set of LEO satellites not represented in training, the mean absolute error against direct OTSO tracing is $0.200$~GV (Figure S11 of the Supporting Information), with the largest residuals concentrated at the lowest rigidities. This out-of-distribution MAE is the value carried into the uncertainty propagation in Section~\ref{method:vulnerability}.

\subsection{Vulnerability assessment}\label{method:vulnerability}

With storm-time cutoff rigidities established for each satellite trajectory in Section~\ref{method:cutoff}, satellite vulnerability to SEP exposure is quantified through a three-stage framework connecting geomagnetic particle access, radiation dose transport, and device failure probability.

Solar energetic protons of kinetic energy $E_k$ access a satellite location when their rigidity exceeds the local cutoff $R_c(t)$. The particle access fraction is defined as:
\begin{equation}
    f_{\mathrm{access}}(>E_k) = \frac{1}{N} \sum_{i=1}^{N} 
    \mathds{1}\left[R(E_k) \geq R_c(t_i) + \delta R\right]
    \label{eq:f_access}
\end{equation}
where $\mathds{1}[\cdot]$ is the indicator function returning $1$ when its argument is true and $0$ otherwise, $N$ is the number of trajectory points, $R(E_k)$ is the rigidity corresponding to kinetic energy $E_k$, and $\delta R = 0.05$~GV is a margin to reduce threshold sensitivity. Access fractions are computed for 10,650 satellites: 311 from OTSO particle tracing (all orbital regimes) and 10,339 from ML inference (predominantly LEO). Uncertainty is propagated through the out-of-distribution $R_c$ prediction error (MAE~$= 0.200$~GV, see Supporting Information Text S5), with reduced uncertainty applied to satellites where $R_c$ comes directly from OTSO physics-based tracing. The effective fluence combines the $T$-year return level fluence from the GPD analysis in Table S8 of the Supporting Information with the access fraction:
\begin{equation}
    \Phi_{\mathrm{eff}}(>E) = \Phi_{\mathrm{ext}}(>E) 
    \cdot f_{\mathrm{access}}(>E)
    \label{eq:phi_eff}
\end{equation}

Converting effective fluence to absorbed dose requires knowledge of the full proton energy spectrum behind the shielding. The five integral fluence return levels from the GPD analysis define the spectral shape through piecewise power-law interpolation:
\begin{equation}
    J_{>E_k} = J_{>E^{(n)}}\left(\frac{E_k}{E^{(n)}}\right)^{-\gamma^{(n)}}, 
    \quad E^{(n)} \leq E_k < E^{(n+1)}
    \label{eq:powerlaw}
\end{equation}
\noindent
where $E^{(n)} \in \{5, 10, 30, 50, 100\}$~MeV are the five tabulated thresholds, and the local spectral index is
\begin{equation}
    \gamma^{(n)} = \frac{\ln\left[J_{>E^{(n)}} / J_{>E^{(n+1)}}\right]}{\ln\left[E^{(n+1)} / E^{(n)}\right]}.
    \label{eq:spectral_index}
\end{equation}

The differential spectrum $j(E) = -dJ_{>E}/dE$ is evaluated at 25 energies spanning 0.1--2000~MeV and transported through aluminum spherical shell shielding using the International Radiation Environment Near Earth (IRENE) radiation transport framework \cite{ginet_ae9_2013}, which applies the SHIELDOSE-2 dose calculation methodology \cite{seltzer_updated_1994} to compute absorbed dose in silicon. Doses are computed at shielding depths of 10--1000~mil~Al for return periods of 50--1200 years, producing a lookup table of $D_{\mathrm{SEP}}$(shielding, $T$) at $f_{\mathrm{access}} = 1$. For a specific satellite, the SEP dose contribution is scaled by its access fraction.

The cumulative TID from trapped radiation is computed using the AP9 and AE9 models \cite{ginet_ae9_2013} within the same IRENE framework. A precomputed grid of (22 altitudes) $\cdot$ (10 inclinations) $\cdot$ (6 shielding depths) provides the baseline TID for a standardized 1-year mission duration at circular orbits. Each satellite's cumulative trapped dose is obtained by interpolating this grid at its mean orbital parameters and scaling by the elapsed years since launch through January 2026. This linear scaling approach assumes that the trapped radiation environment remains statistically stationary over mission lifetimes, which is reasonable given that the AP9 and AE9 models represent long-term statistical averages that inherently account for solar cycle variations and historical SEP contributions to the trapped particle population. The total dose environment for a given satellite is:
\begin{equation}
    D_{\mathrm{total}} = D_{\mathrm{trapped}}
    (\mathrm{alt},\,\mathrm{inc},\,\mathrm{years},\,\mathrm{shield}) 
    + f_{\mathrm{access}} \cdot D_{\mathrm{SEP}}
    (\mathrm{shield},\,T)
    \label{eq:tid_total}
\end{equation}

Since satellite-specific shielding and component data are generally proprietary, regime-dependent assumptions are adopted for shielding thickness and device radiation hardness, as summarized in Table~\ref{tab:regime_assumptions}. These values represent central estimates for each regime and carry substantial uncertainty, as shielding thickness, component selection, and radiation hardness are highly mission-dependent and vary considerably even within a single orbital regime. LEO constellations predominantly use commercial off-the-shelf (COTS) components behind moderate shielding, while MEO navigation and GEO telecommunications satellites employ radiation-hardened or radiation-tolerant parts with heavier shielding. HEO missions, in particular, span a wide range of designs depending on orbit geometry, from highly robust 12-hour Molniya-type orbits that traverse both radiation belts to Tundra-type orbits with lower accumulated dose, more comparable to GEO.

\begin{table}[H]
\centering
\caption{\textbf{Regime-dependent shielding and device failure assumptions.} Shielding thicknesses reflect central estimates of typical design practices for each orbital regime. Actual values are highly mission-dependent and carry substantial uncertainty. Device failure dose parameters are lognormal ($\mu$: median failure dose, $\sigma$: log-space standard deviation), following the methodology of \citeA{xapsos_inclusion_2017}.}
\label{tab:regime_assumptions}
\vspace{2mm}
\footnotesize
\setlength{\tabcolsep}{4pt}
\renewcommand{\arraystretch}{1.0}
\begin{tabular*}{0.9\linewidth}{@{}l @{\extracolsep{\fill}} cccc@{}}
\hline
\textbf{Regime} & \textbf{Shielding (mil~Al)} & \textbf{Component class} & \textbf{$\mu$ (krad)} & \textbf{$\sigma$} \\
\hline
LEO & 100 & Commercial/COTS & 10 & 0.6 \\
MEO & 500 & Radiation-hardened & 1000 & 0.3 \\
GEO & 300 & Radiation-tolerant & 500 & 0.3 \\
HEO & 200 & Mixed heritage & 20 & 0.5 \\
\hline
\end{tabular*}
\end{table}

The probability that a device fails due to total dose follows the framework of \citeA{xapsos_inclusion_2017}:
\begin{equation}
    P_{\mathrm{fail}} = \int_0^{\infty} 
    \left[1 - H(x)\right] g(x)\,dx,
    \label{eq:pfail}
\end{equation}
where $H(x)$ is the CDF of the total dose environment and $g(x)$ is the lognormal probability density function (PDF) of device failure doses, as specified in Table~\ref{tab:regime_assumptions}.

The environment distribution $H(x)$ is constructed by propagating uncertainties in shielding thickness and $f_{\mathrm{access}}$ through the dose chain via Monte Carlo sampling ($n = 500$ per satellite). The resulting $P_{\mathrm{fail}}$ provides a physically grounded vulnerability metric that accounts for orbital environment, shielding, accumulated trapped dose, and device hardness simultaneously. Satellites are classified into five vulnerability categories based on $P_{\mathrm{fail}}$, as defined in Table~\ref{tab:vulnerability_class}.

\begin{table}[H]
\centering
\caption{\textbf{Vulnerability classification based on failure probability.} Categories reflect the probability that a randomly selected device on the satellite fails due to total ionizing dose during a 1-in-100-year SEP event, given regime-dependent shielding and component assumptions (Table~\ref{tab:regime_assumptions}).}
\label{tab:vulnerability_class}
\vspace{2mm}
\footnotesize
\setlength{\tabcolsep}{4pt}
\renewcommand{\arraystretch}{1.0}
\begin{tabular*}{0.9\linewidth}{@{}l @{\extracolsep{\fill}} cl@{}}
\hline
\textbf{Class} & \textbf{$P_{\mathrm{fail}}$} & \textbf{Interpretation} \\
\hline
Critical & $> 10^{-2}$           & High likelihood of dose-induced failure \\
Elevated & $10^{-3} - 10^{-2}$   & Significant risk, margins likely insufficient \\
Moderate & $10^{-6} - 10^{-3}$   & Within typical design margin uncertainty \\
Low      & $10^{-9} - 10^{-6}$   & Well within design margins \\
Negligible & $< 10^{-9}$         & Below the operational resolution of the classification \\
\hline
\end{tabular*}
\end{table}

\subsection{Economic impact assessment}\label{method:economic}

The translation of SEP vulnerability into economic impact requires linking failure probabilities to satellite service disruptions and, subsequently, to sectoral economic losses.

\subsubsection{Satellite cost model}

Satellite replacement values are estimated using a three-tier hybrid approach. For mega-constellations and government programs with documented costs, unit values are assigned directly from public sources (Tier 1). Starlink satellites are valued at \$0.25--0.80M based on industry analyst estimates \cite{erwin_starlink_2024}. Navigation satellites command significantly higher values: GPS III at \$250--340M based on Space Force contract awards \cite{erwin_gps_iiif_2025, govconwire_gps_iiif_2022}. Military communications satellites, including Advanced Extremely High Frequency (AEHF), are valued at \$850M--1.8B per unit, with total program costs of \$15.5B for six satellites \cite{erwin_aehf6_2020, erwin_military_space_costs_2018}. Space-Based Infrared System (SBIRS) satellites are valued at approximately \$930M based on the \$1.86B contract for GEO-5 and GEO-6 \cite{afcea_sbirs_2014}. Weather satellites GOES-R series cost approximately \$1.9B per unit based on a \$7.69B program for four satellites \cite{gao_goes_r_2008, spacenews_goes_r_2007}. Joint Polar Satellite System (JPSS) satellites cost approximately \$3.8B based on the \$18.8B life-cycle cost for five satellites \cite{leone_jpss_2012, physics_today_jpss_2012}.

For satellites outside known constellations, a parametric model estimates cost based on mission classification and orbital regime (Tier 2). Satellites lacking either constellation membership or a recognized mission classification fall back to a regime-only default cost range (Tier 3). Both tiers share the same mass and production-volume adjustments, with full details in Supporting Information Text S6. The expected loss for each satellite is computed as:
\begin{equation}
    L_i = C_i \cdot P_{\mathrm{fail},i}
    \label{eq:expected_loss}
\end{equation}
where $C_i$ is the replacement cost and $P_{\mathrm{fail},i}$ is the failure probability from Equation~\ref{eq:pfail}.

\subsubsection{Service-level disruption}

Economic impact is assessed through three failure scenarios that progressively expand the set of satellites assumed to fail permanently during a 1-in-100-year SEP event, as outlined in Table~\ref{tab:scenarios}. For each scenario, failed satellites lose 100\% of their service capacity, and the disruption level for each service category is the fraction of service value lost:
\begin{equation}
    D_j = \frac{\sum_{i \in \mathrm{failed}_j} C_i}{\sum_{i \in \mathrm{service}_j} C_i}
    \label{eq:disruption}
\end{equation}
where $D_j$ is the disruption fraction for service $j$, the numerator sums the replacement costs of failed satellites in service $j$, and the denominator is the total value of all satellites in that service. Satellites are assigned to five service categories (navigation, communications, weather, Earth observation (EO), military/intelligence) based on constellation membership or mission classification.

\begin{table}[H]
\centering
\caption{\textbf{Failure scenarios for economic impact assessment.} Each scenario defines a $P_{\mathrm{fail}}$ threshold above which satellites are assumed to fail permanently during the 1-in-100-year SEP event. Scenarios correspond to the vulnerability classification in Table~\ref{tab:vulnerability_class}.}
\label{tab:scenarios}
\vspace{2mm}
\footnotesize
\setlength{\tabcolsep}{4pt}
\renewcommand{\arraystretch}{1.0}
\begin{tabular*}{0.9\linewidth}{@{}l @{\extracolsep{\fill}} lc@{}}
\hline
\textbf{Case} & \textbf{$P_{\mathrm{fail}}$ threshold} & \textbf{Description} \\
\hline
Case 1 (likely)      & $> 10^{-2}$ & Critical satellites fail \\
Case 2 (moderate)    & $> 10^{-3}$ & Critical and elevated satellites fail \\
Case 3 (worst case)  & $> 10^{-6}$ & All satellites with non-negligible risk \\
\hline
\end{tabular*}
\end{table}

\subsubsection{Input-output framework}

Sectoral losses are computed using input-output analysis to capture inter-industry propagation effects \cite{miller_inputoutput_2009}. A sector-service dependency matrix $\delta_{sj}$ specifies the fraction of each economic sector $s$'s productivity dependent on satellite service category $j$ (see Supporting Information Text S7). The total sectoral shock combines the service disruptions weighted by dependencies:
\begin{equation}
    \alpha_s = \sum_{j \in \mathrm{services}} \delta_{sj} \cdot D_j
    \label{eq:sector_shock}
\end{equation}
\noindent
where $\alpha_s$ is the total productivity shock to sector $s$. Direct value-added loss follows as $\Delta V_s = \alpha_s \cdot V_s^0 / 365$ where $V_s^0$ is the annual sectoral value-added and $\Delta V_s$ is the daily direct loss.

Total output impact, including indirect effects, is computed via the Ghosh supply-driven inverse \cite{ghosh_inputoutput_1958, dietzenbacher_vindication_1997}:
\begin{equation}
    \Delta X_s = \sum_r \Delta V_r \cdot G_{rs}
    \label{eq:ghosh}
\end{equation}
\noindent 
where $\Delta X_s$ is the total output loss for sector $s$, $\Delta V_r$ is the direct value-added loss for sector $r$, $\mathbf{G} = (\mathbf{I} - \mathbf{B})^{-1}$ is the Ghosh inverse matrix capturing forward linkages through supply chains, $G_{rs}$ is the $(r,s)$ element of $\mathbf{G}$, $\mathbf{I}$ is the identity matrix, and $\mathbf{B}$ is the direct output coefficient matrix. The analysis employs a 13-sector aggregation of the Bureau of Economic Analysis (BEA) Use Tables \cite{bea_inputoutput_2023}.

Uncertainty is propagated via Monte Carlo simulation ($N_{io} = 10{,}000$ draws) over the sector-service dependency parameters, which are specified as distributions (mean $\pm$ standard deviation) rather than point estimates (see Supporting Information Text S9). Service disruption levels are deterministic within each scenario since the failure criterion is binary.

\newpage
\section{Results}\label{sec:results}
\subsection{Vulnerability assessment}

\begin{figure}[H]
    \centering
    \includegraphics[width=\textwidth]{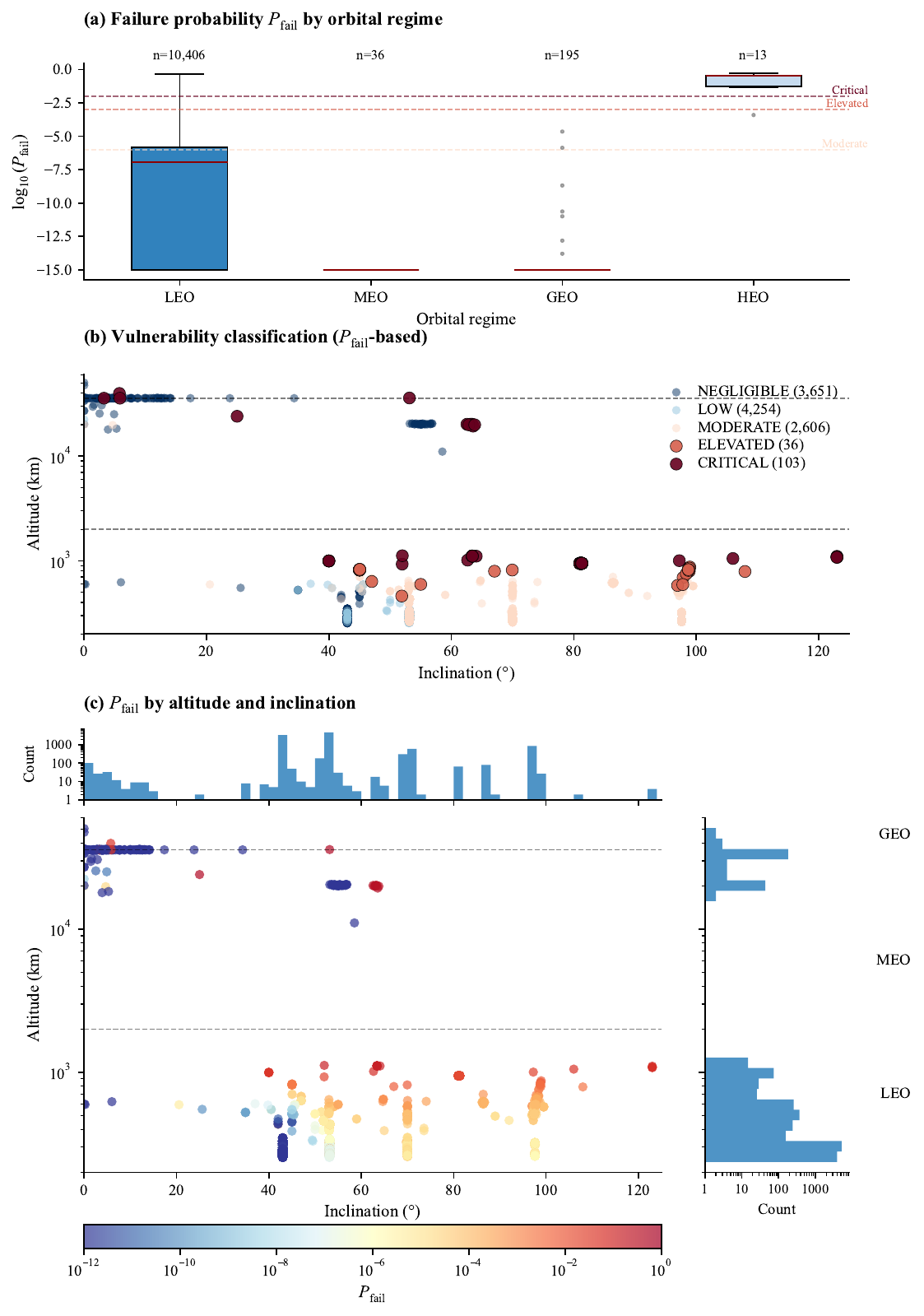}
    \vspace{-0.6in}
    \caption{\textbf{SEP vulnerability assessment of the operational satellite fleet.} (a) Distribution of $P_{\mathrm{fail}}$ by orbital regime, with classification thresholds indicated. (b) Vulnerability classification mapped onto altitude-inclination space. Critical and Elevated risk satellites concentrate in high-altitude LEO ($>$1000~km) at high inclinations and in HEO. (c) Continuous $P_{\mathrm{fail}}$ values showing the transition from negligible risk at low altitudes and inclinations to elevated risk in the inner proton belt and polar orbits.}
    \label{fig:vulnerability}
\end{figure}

The failure probability analysis reveals a strongly heterogeneous risk distribution across the 10,650 satellites assessed, as shown in Figure~\ref{fig:vulnerability}. Mean particle access fractions span more than an order of magnitude across regimes, from $\bar{f}_{\mathrm{access}}(>30~\mathrm{MeV}) = 0.04$ in LEO to 0.44 in HEO, with the full breakdown by energy threshold given in Table S3. Under the 1-in-100-year SEP scenario with regime-dependent shielding and the hardness assurance methodology of \citeA{xapsos_inclusion_2017}, as detailed in Table~\ref{tab:regime_assumptions}, 103 satellites (1.0\%) are classified as Critical ($P_{\mathrm{fail}} > 10^{-2}$), 36 as Elevated risk ($10^{-3} < P_{\mathrm{fail}} \leq 10^{-2}$), and 2,606 as Moderate ($10^{-6} < P_{\mathrm{fail}} \leq 10^{-3}$). The remaining 7,905 satellites (74.2\%) fall into Low or Negligible categories, with failure probabilities below $10^{-6}$.

\newpage
\newpage
The orbital regime strongly determines vulnerability, but through mechanisms distinct from geomagnetic access alone. The mean trapped-radiation baseline dominates the dose budget: LEO satellites accumulate a mean TID of 0.181~krad(Si) per year behind 100~mil~Al shielding, while the 1-in-100-year SEP event contributes only 0.012~krad(Si) on average. The most vulnerable LEO satellites orbit at altitudes above 1000~km and at inclinations exceeding $50^\circ$, where they reside within the inner proton belt and accumulate baseline doses approaching or exceeding the assumed 10~krad commercial component failure threshold within a few years of operation. For these satellites, the extreme SEP event represents an additional stress that pushes already marginal dose budgets past the failure point.

HEO satellites exhibit the highest mean $P_{\mathrm{fail}}$ of 0.27, with 12 of 13 assets classified as Critical. These satellites repeatedly traverse the radiation belts and accumulate dose rapidly despite moderate shielding (200~mil~Al). In contrast, under the assumed shielding and hardness levels, MEO and GEO satellites fall in the Negligible class ($P_{\mathrm{fail}} < 10^{-9}$, Table~\ref{tab:vulnerability_class}) owing to radiation-hardened or radiation-tolerant components and heavy shielding that provide large design margins relative to both trapped and SEP dose environments. The lognormal convolution yields smaller numerical values that fall outside the calibration range of the device hardness data and are not interpreted further.

\subsection{Fleet cost and expected loss}

\begin{figure}[H]
    \centering
    \includegraphics[width=\textwidth]{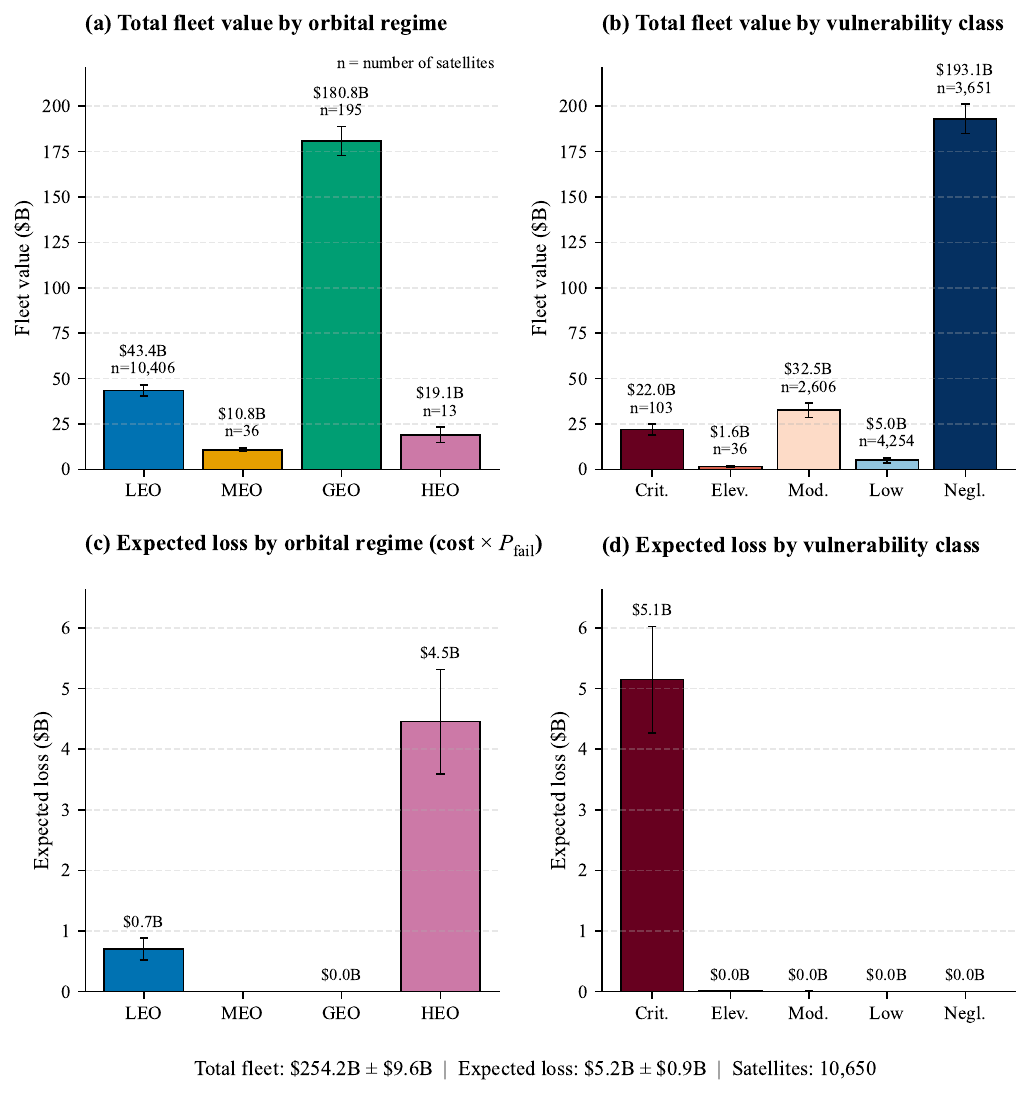}
    \caption{\textbf{Fleet value and expected loss assessment.} (a) Total fleet value by orbital regime. (b) Fleet value by $P_{\mathrm{fail}}$-based vulnerability classification. (c) Expected loss (cost $\times$ $P_{\mathrm{fail}}$) by regime. (d) Expected loss by vulnerability class. Error bars represent propagated uncertainty from cost estimation. The Negligible class holds \$193.1B in value but contributes zero expected loss, demonstrating the effectiveness of radiation-hardened design for GEO and MEO assets under the assumed shielding.}
    \label{fig:costs}
\end{figure}

The satellite fleet represents a replacement value of \$254.18B $\pm$ \$9.6B, as illustrated in Figure~\ref{fig:costs}. GEO assets account for the largest share (\$180.8B, 71\%) despite comprising only 1.8\% of satellites by count, reflecting the concentration of high-value communications, weather, and military platforms in geosynchronous orbit. LEO satellites total \$43.4B in value, dominated by mega-constellation assets.

The expected loss, computed as replacement cost weighted by failure probability, totals \$5.16B $\pm$ \$0.88B across the fleet. This exposure is concentrated almost entirely in two categories: HEO assets contribute \$4.46B (86\%) from only 13 satellites, reflecting the combination of high individual asset values and high failure probabilities. LEO contributes \$0.70B (14\%) spread across 91 Critical satellites. MEO and GEO contribute negligible expected loss despite holding \$191.7B in combined fleet value, demonstrating that radiation-hardened design and heavy shielding effectively eliminate dose-induced failure risk for these regimes. At the vulnerability-class level, the 103 Critical satellites carry a replacement value of \$22.0B, and the 36 Elevated risk satellites a further \$1.6B, for a combined \$23.6B exposed to potential dose-induced failure under the modeled scenario.

The vulnerability-value relationship exhibits a striking asymmetry. The 103 Critical satellites represent 1.0\% of the fleet by count but capture 99\% of the expected loss (\$5.16B). The 3,651 Negligible satellites hold 76\% of fleet value (\$193.1B) but contribute zero expected loss.

\subsection{Economic impact scenarios}

\begin{figure}[H]
    \centering
    \includegraphics[width=\textwidth]{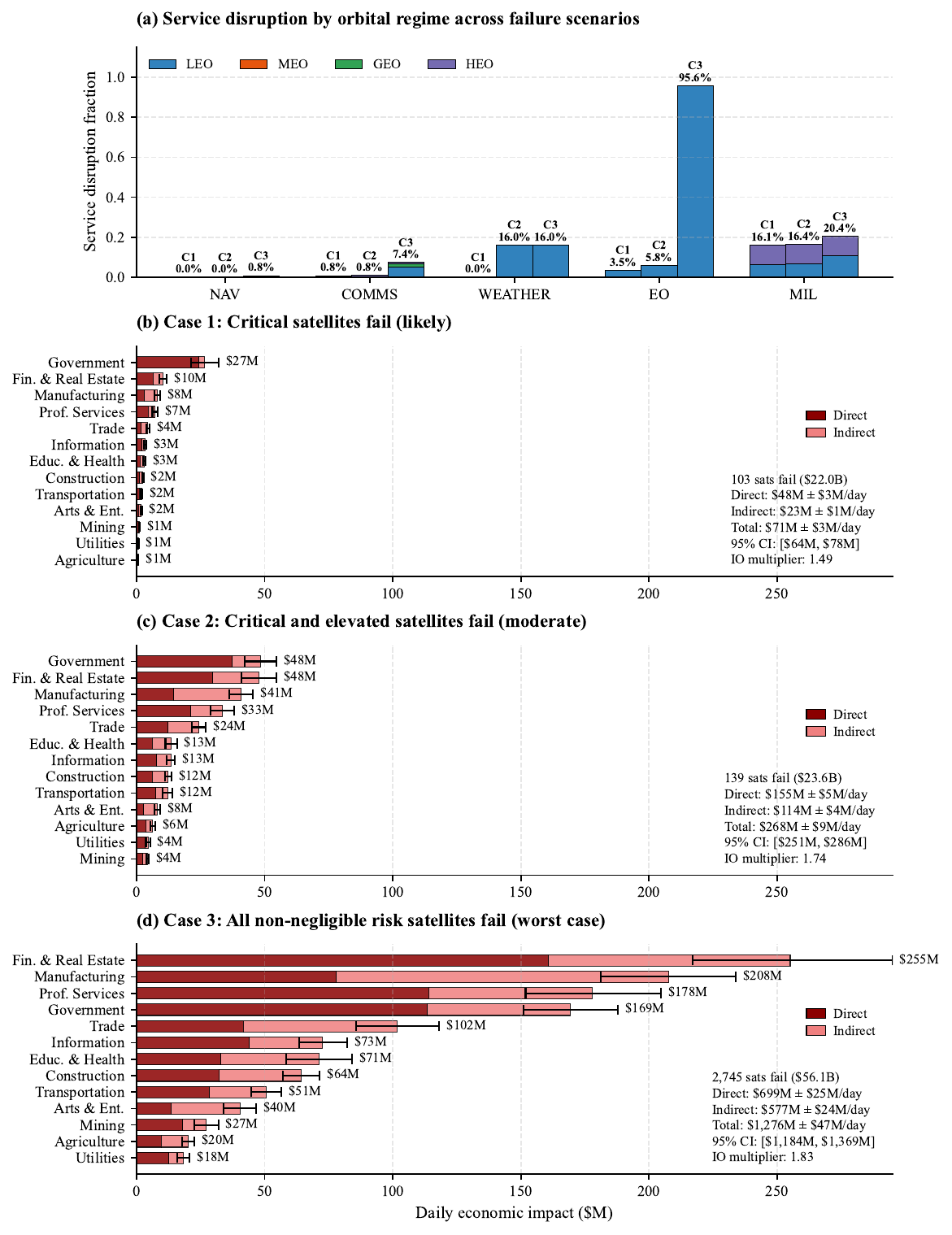}
    \caption{\textbf{Economic impact under three failure scenarios.} (a) Service disruption fraction by orbital regime across scenarios (Case 1--Case 3). Earth observation is the most vulnerable service, with 95.6\% capacity loss under Case 3, driven by polar LEO satellites with commercial components. Military services show 16.1--20.4\% disruption across scenarios, concentrated in HEO early warning assets. (b--d) Daily sectoral economic impact decomposed into direct losses and indirect supply-chain effects for each scenario. Error bars indicate 95\% confidence intervals from a Monte Carlo simulation over sector-service dependency parameters.}
    \label{fig:econ}
\end{figure}

Three failure scenarios translate satellite-level $P_{\mathrm{fail}}$ into service disruption and daily economic impact, as shown in Figure~\ref{fig:econ}.

Under Case 1 (likely), 103 Critical satellites fail, producing \$71.1M $\pm$ \$3.4M in daily economic impact with an IO multiplier of 1.49. Military services experience the highest disruption (16.1\%), driven by HEO early warning and intelligence assets. Government and finance sectors absorb the largest absolute impacts (\$26.6M and \$10.2M per day, respectively).

Case 2 (moderate) adds 36 Elevated risk satellites, expanding failures to 139 and increasing daily impact to \$268.5M $\pm$ \$8.9M. The critical change is the appearance of weather service disruption (16.0\%) as two JPSS polar weather satellites cross the failure threshold, along with substantial EO degradation (5.8\%). The IO multiplier increases to 1.74, reflecting the broader sectoral dependencies activated by weather and EO service losses.

Case 3 (worst case) extends to 2,745 satellites with $P_{\mathrm{fail}} > 10^{-6}$, yielding \$1.28B $\pm$ \$47M in daily impact. EO suffers near-total disruption (95.6\%) as the majority of polar LEO satellites with commercial components fail. Finance and real estate experience the largest absolute impact (\$255.1M/day), followed by manufacturing (\$207.7M/day) and professional services (\$177.7M/day). The IO multiplier reaches 1.83, indicating that supply-chain amplification accounts for 45\% of total losses.

Across all scenarios, service disruption is driven almost entirely by LEO and HEO failures. No MEO or GEO satellites fail in any scenario under the assumed shielding and component hardness, confirming that radiation-hardened components behind heavy shielding provide adequate margins even for century-scale extreme events. The escalation from Case 1 to Case 3 is nonlinear: a 27-fold increase in failed satellites produces an 18-fold increase in economic impact, reflecting the inclusion of progressively lower-value but more numerous LEO assets.

\newpage
\section{Discussion}\label{sec:discussion}

The results reveal a highly stratified vulnerability landscape across the operational satellite fleet, with failure risk concentrated in high-altitude LEO and HEO assets rather than the high-value GEO platforms that dominate fleet replacement cost.

\subsection{What is the SEP exposure of US satellites during extreme storms?}

The cutoff rigidity analysis confirms that the orbital regime is the primary determinant of SEP exposure. During the 1-in-100-year storm scenario, storm-time magnetospheric compression drives substantial cutoff suppression, with $R_c$ dropping by 1--1.5~GV at mid-latitudes. As reported in Table S3 of the Supporting Information, GEO, MEO, and HEO assets all exhibit elevated mean access fractions at $>$30~MeV of 0.40, 0.36, and 0.44, respectively. These orbits lie beyond effective geomagnetic shielding during disturbed conditions. LEO satellites benefit substantially from residual shielding ($\bar{f}_{\mathrm{access}}(>30~\mathrm{MeV}) = 0.04$ on average), though high-inclination and polar-orbiting assets experience elevated exposure where cutoff rigidities fall below the 30~MeV proton threshold.

However, exposure alone does not determine vulnerability. The dose transport analysis demonstrates that the translation from particle access to failure probability depends critically on three additional factors: accumulated trapped radiation dose, shielding thickness, and component radiation hardness. GEO satellites, despite high SEP access, fall in the Negligible class ($P_{\mathrm{fail}} < 10^{-9}$) under the assumed shielding (300~mil~Al) and component hardness (500~krad), because the total dose environment, including both trapped and SEP contributions, sits well within design margins.

\subsection{What is the vulnerability of the US operational satellite fleet?}

The $P_{\mathrm{fail}}$-based assessment identifies two distinct populations of vulnerable satellites. The first comprises high-altitude LEO satellites (above 1000~km) at inclinations exceeding $50^\circ$, which reside within the inner proton belt and accumulate trapped-radiation doses that approach commercial component failure thresholds within a few years of operation. For these 91 LEO satellites classified as Critical, the 1-in-100-year SEP event acts as an additional dose increment that pushes already marginal radiation budgets past the failure point. The second comprises 12 HEO assets that repeatedly traverse the full depth of the radiation belts, accumulating dose rapidly despite moderate shielding.

The expected loss of \$5.16B $\pm$ \$0.88B is concentrated almost entirely in these two populations: HEO contributes \$4.46B (86\%) from only 13 satellites, while LEO contributes \$0.70B (14\%) across 91 Critical satellites. MEO and GEO contribute negligible expected loss despite holding \$191.7B in combined fleet value.

Several factors moderate the practical implications of these results. Navigation services exhibit inherent resilience through constellation architecture: GPS requires a minimum of 24 operational satellites for global coverage, with 31 currently active and additional spares on orbit \cite{us_department_of_defense_global_2020}. No GPS satellites fail in Case 1 or Case 2 under the assumed radiation-hardened design (1000~krad failure threshold behind 500~mil~Al), and only two GPS satellites cross the failure threshold in Case 3 (0.8\% disruption), well within the constellation's 24-satellite minimum. Communications services, dominated by Starlink's satellites at low altitude and low inclination, show minimal disruption (0.8--7.4\% across scenarios) due to substantial geomagnetic protection and, as noted by \citeA{guo_resilience_2025}, mesh network topology that enables dynamic rerouting around failed nodes.

Weather and EO services present distinct vulnerabilities. Weather satellites show disruption only in Case 2 and beyond (16\%), when two JPSS polar weather satellites cross the failure threshold. The small constellation size and lack of operational redundancy make this a disproportionate risk. EO suffers the most severe degradation, reaching 95.6\% disruption under Case 3 as the majority of polar LEO satellites with commercial components fail. This reflects the structural vulnerability of a service category that depends predominantly on COTS-based satellites in high-inclination orbits, precisely the orbital regime where trapped-proton-belt doses are highest.

Military and intelligence assets show 16.1--20.4\% disruption across scenarios, concentrated in HEO early warning and intelligence, surveillance, and reconnaissance (ISR) platforms. The GEO communications backbone, including AEHF, Wideband Global SATCOM (WGS), and Milstar, falls in the Negligible class under the assumed radiation-hardened components and 300~mil~Al shielding, so no GEO communications satellite crosses any of the Case 1 through Case 3 failure thresholds. Operators can mitigate risk by transitioning to safe mode during SEP events \cite{blumenfeld_its_2024}, though this introduces service interruptions distinct from permanent hardware failure.

\subsection{What are the economic consequences of SEP-induced disruptions?}

The scenario-based economic impact assessment yields daily losses ranging from \$71.1M (Case 1, likely) to \$1.28B (Case 3, worst case), with the moderate scenario (Case 2) producing \$268.5M per day. The IO multiplier increases from 1.49 to 1.83 across scenarios, reflecting the broader sectoral dependencies activated as more service categories experience disruption. The dominant sectors shift across scenarios. In Case 1, government services absorb the largest absolute impact (\$27M per day), driven by failures of HEO early warning and intelligence, surveillance, and reconnaissance assets. In Case 2, the government and the finance and real estate sectors are roughly co-equal at \$48M per day each, as weather and Earth observation losses begin to engage broader supply chains. In Case 3, finance and real estate dominate at \$255M per day, ahead of manufacturing (\$208M), professional services (\$178M), and government (\$169M), reflecting the range of navigation and communications dependencies once the largest LEO populations fail.

The escalation from Case 1 to Case 3 is nonlinear: a 27-fold increase in failed satellites produces an 18-fold increase in economic impact, reflecting the inclusion of progressively lower-value but more numerous LEO assets. The transition from Case 1 to Case 2 is particularly significant, as it introduces weather service disruption and substantially increases EO degradation, activating sectoral dependencies (agriculture, transportation) that are not engaged under Case 1.

These estimates should be interpreted as upper bounds on daily impact for several reasons. First, the analysis assumes permanent satellite failure, whereas operational responses, including safe mode, attitude adjustments, and deferred operations, can preserve hardware at the cost of a temporary service interruption. Second, economic actors possess adaptation capacity through terrestrial alternatives and operational workarounds. Third, productivity losses during storm periods may be partially recoverable through compensatory effort in subsequent days, implying that daily impact estimates overstate cumulative welfare losses.

The scenario framework itself represents a methodological advance over single-point estimates. By linking failure probability thresholds to economic impact through a physically grounded dose transport chain, the analysis provides decision-makers with a range of outcomes tied to explicit assumptions about which satellites fail, rather than a single number that obscures the underlying uncertainty.
\newpage

\section{Conclusion}\label{sec:conclusion}

This study developed an integrated framework linking SEP hazard characterization, dynamic cutoff-rigidity modeling, radiation-dose transport, and fleet-wide failure probability estimation to quantify satellite vulnerability and the economic consequences under extreme space weather scenarios.

The analysis of 10,650 operational satellites reveals that failure risk is concentrated in high-altitude LEO and HEO orbits, where trapped proton belt doses combined with contributions from SEP events can exceed commercial component failure thresholds. Of the fleet, 103 satellites (1.0\%) are classified as Critical ($P_{\mathrm{fail}} > 10^{-2}$) and an additional 36 as Elevated risk, together representing a total of \$23.6B in replacement value. In contrast, MEO navigation and GEO telecommunications satellites fall into the Negligible class ($P_{\mathrm{fail}} < 10^{-9}$) under the assumed radiation-hardened design, despite high particle access fractions during storm conditions.

The expected loss of \$5.16B and daily economic impact ranging from \$71M to \$1.28B across three failure scenarios demonstrate that the economic consequences of extreme SEP events are substantial but highly concentrated in HEO and high-altitude LEO assets.

The key methodological contribution is the integration of probabilistic SEP fluence estimation (GPD with method of moments), spectral dose reconstruction through SHIELDOSE-2 transport, trapped radiation baseline from AP9/AE9, and the \citeA{xapsos_inclusion_2017} failure probability framework into a single pipeline applicable to fleet-scale assessment. By replacing the traditional radiation design margin approach with physics-based $P_{\mathrm{fail}}$ and linking it to scenario-based economic impact through input-output analysis, the framework provides actionable information for fleet operators, insurers, and policymakers.

Several limitations constrain interpretation. Most fundamentally, the analysis considers only total ionizing dose. Transient single-event effects, including single-event upsets and latch-ups triggered by individual energetic protons, are not modeled and could produce additional service interruptions during the SEP event, independent of the dose-induced failures captured here. The analysis assumes regime-dependent shielding and component hardness (see Table~\ref{tab:regime_assumptions}) in lieu of satellite-specific design data, which is generally proprietary. The SEP dose contribution is computed from a piecewise power-law spectral reconstruction, validated against monoenergetic estimates using proton stopping powers in silicon from \citeA{berger_stopping_2017}, yielding IRENE-to-Berger dose ratios of 0.51--1.31 across shielding depths, which introduces spectral shape uncertainty. Moreover, the SEP fluence reaching each satellite is treated as omnidirectional throughout the storm window, with the access fraction $f_{\mathrm{access}}$ capturing only the geomagnetic cutoff filtering. The geometric eclipse of the satellite by Earth's shadow, which removes line of sight to the SEP source for roughly 30--40\% of each orbit in circular LEO below 1000~km, is not modeled and contributes to overestimating exposure for low-altitude satellites. Particle anisotropy along the interplanetary magnetic field is similarly neglected, although individual events frequently show pronounced field-aligned distributions that would change the effective fluence at any given orbital location. The cost model relies on publicly available contract values, and the sector-service dependency matrix reflects expert judgment rather than empirical measurement. The analysis treats the fleet as static, whereas operators receive 1--2 days' warning of CME arrival and implement protective measures \cite{blumenfeld_its_2024}. Future research should extend this framework to incorporate satellite-specific shielding data where available, multi-event cumulative effects across solar cycles, and operator response protocols that trade temporary service degradation for hardware preservation.

These findings directly address Objective 1 (infrastructure protection) and advance Objective 2 (characterization and forecasting) of the National Space Weather Strategy \cite{ostp_national_2019} through the integrated SEP hazard-to-impact pipeline. The framework aligns with the 2023 Implementation Plan \cite{sworm_implementation_2023}, which assigns benchmarks, hazard mapping, and impact-based products that translate strategy into operations.

\acknowledgments
This material is based upon work supported by the NSF National Center for Atmospheric Research, which is a major facility sponsored by the US National Science Foundation under Cooperative Agreement No. $1852977$. The project upon which this article is based was funded through the NSF NCAR Early-Career Faculty Innovator Program under the same Cooperative Agreement. We also gratefully acknowledge funding from the ChronoStorm NSF RAPID grant (\#$2434136$), co-funded by the GEO/AGS Space Weather Research and the ENG/CMMI Humans, Disasters, and the Built Environment programs. 

We also thank Paul O'Brien (The Aerospace Corporation) for reading an earlier version of this manuscript and for his invaluable contributions to the radiation environment methodology.

\section*{Open Research Section}
Python is used for data collection, modeling, and analysis. Cutoff rigidities are computed with the Oulu Open-source geomagneToSphere propagation tool (OTSO) accessed via the OTSOpy interface \cite{larsen_otso_2023}. Trapped radiation environments and SEP dose transport are computed with the International Radiation Environment Near Earth (IRENE) framework, which implements the AE9/AP9 climatological models \cite{ginet_ae9_2013} and the SHIELDOSE-2 dose calculation methodology \cite{seltzer_updated_1994}. Satellite orbital elements were obtained from Space-Track (\url{https://www.space-track.org}) and supplemented with catalog data from Jonathan's Space Report (\url{https://www.planet4589.org}). The Solar Energetic Particle Environment Modelling (SEPEM) Reference Data Set v3.2 \cite{crosby_sepem_2015} provides quality-controlled integral proton flux for 1974--2017. GOES-16 Solar and Galactic Proton Sensor data are obtained from the National Oceanic and Atmospheric Administration (NOAA). Interplanetary and geomagnetic parameters are obtained from the NASA/GSFC OMNI database \cite{nasa_omniweb}. Proton stopping powers in silicon and aluminum are obtained from the NIST PSTAR database \cite{berger_stopping_2017}. The project code, analysis pipelines, and visualizations are available on GitHub (\url{https://github.com/denniesbor/C-SWIMs}). Input data, trained model weights, and intermediate pipeline outputs are archived on Zenodo \cite{oughton_cswims_2026}.

\section*{Conflict of Interest}
The authors declare that this research was conducted in the absence of any commercial or financial relationships that could be construed as a potential conflict of interest.

\newpage
\bibliography{references}
\end{document}